\documentclass{ws-procs9x6}

\begin{document}

\title{Grand unification of flavor by orbifold twisting $Z_2$ and
$Z_2\times Z_2^\prime$}

\author{Jihn E. Kim}

\address{School of Physics, \\
Seoul National University, \\
Seoul 151-747, Korea\\
E-mail: jekim@phyp.snu.ac.kr}


\maketitle

\abstracts{The grand unification of flavor(GUF) in extra
dimensions is discussed. After reviewing the old GUF, I present a
GUF model $SU(7)$ in 5D with the recently popular field theoretic
orbifold compactification.}

\newcommand{\bea}{\begin{eqnarray}}
\newcommand{\eea}{\end{eqnarray}}
\def\beq{\begin{equation}}
\def\eeq{\end{equation}}

\def\one{\bf 1}
\def\two{\bf 2}
\def\five{\bf 5}
\def\ten{\bf 10}
\def\tenb{\overline{\bf 10}}
\def\fiveb{\overline{\bf 5}}
\def\threeb{{\bf\overline{3}}}
\def\three{{\bf 3}}
\def\fb{{\overline{F}\,}}
\def\hb{{\overline{h}}}
\def\Hb{{\overline{H}\,}}

\def\slash#1{#1\!\!\!\!\!\!/}

\noindent {\bf Gauge coupling unification}--The simplest(rank 4)
GUT $SU(5)$ classifies the 15 chiral fields in two $SU(5)$
representations, {\bf 10} plus $ \overline{\bf 5}$. Of course, one
can introduce $SU(5)$ singlet(s) which must be electrically
neutral, and guarantees the GUT value of the weak mixing angle
3/8. The next simplest GUT is a rank 5 group $SO(10)$. Its rank 5
subgroup $SU(5)\times U(1)$ can be a GUT group(the so-called {\it
flipped}-SU(5)) with the 16 chiral fields assigned to {\bf 10},
$\overline{\bf 5}$, and {\bf 1}, with {\bf 1} carrying one unit of
electric charge\cite{su51}, which are represented as
\begin{eqnarray}
10_F=\left(\matrix{0\ \ \ d^c\ -d^c\ u\ \ \ d\cr -d^c\ \ 0\ \ \
d^c\ \ \ u\ \ \ d  \cr d^c\ -d^c\ \ \ 0\ \ \ u\ \ \ d\cr -u\ -u\
-u\ \ \ 0\ \ \ \nu_e^c\cr -d\ -d\ -d\ -\nu_e^c\ 0 \cr}\right)_L\
,\ \ \bar 5_F=\left(\matrix{u^c\cr u^c \cr u^c \cr \nu_e\cr
-e\cr}\right)_L\ ,\ \ e^+_L.
\end{eqnarray}
Because, the electromagnetic charge contains the SU(5) singlet
$U(1)$ generator, one can introduce singlets with an arbitrary
electric charge and hence the GUT value of the weak mixing angle
is not predicted, in general. On the other hand, one does not
necessarily need an adjoint representation for breaking the GUT
group since {\bf 10} contains a $SU(3)\times SU(2)\times U(1)$
singlet neutral component $\langle {\bf 10}_{45}\rangle$ which can
be a GUT scale. The electromagnetic charge $Q_{em}$ in the
flipped-$SU(5)$ is
$$
Y=-\frac{1}{15} Y_5+\frac15 Y_0.
$$
Since the flipped $SU(5)$ is a subgroup of $SO(10)$, $-Y_0(1)=
5Y_0(\bar 5)+10Y_0(10)$, predicting the hypercharges in the ratio
--1: 3: --5 for {\bf 10}, $\overline{\bf 5}$, and {\bf 1},
respectively. This flipped-$SU(5)$ was discussed in ordinary
GUTs\cite{su51}, SUSY GUTs\cite{dkn}, and string GUTs\cite{nahe}.

But the flipped $SU(5)$ does not realize the unification of the
coupling constants as neatly as in the original $SU(5)$, since
there can exist charged $SU(5)$ singlet members rendering the bare
value of $\sin^2\theta^0\ne 3/8$.
\\

\noindent{\bf Question of flavor}-- The question on the flavor
started by I. I. Rabi by asking, $\lq\lq$Who ordered muon?" Now,
we know that we need at least three generations to have our
universe. With the three family structure of the standard model,
horizontal gauge symmetries such as $SO(3), SU(3)$, etc, have been
considered. But the horizontal symmetry models must be unified if
one insists on the unification of coupling constants. In this
respect, the grand unification of flavor(or families)(GUF) must
replace GUT as emphasized by Georgi\cite{georgi}. Georgi
considered an $SU(11)$ model for three families.

The GUF must be bigger than $SU(5)$. $SU(6)$ cannot have a family
structure. $SU(7)$ can have a family structure. Orthogonal groups
in 4D are the mostly probable GUT or GUF since they automatically
leads to anomaly-free models. Thus, complex spinor represntations
of $SO(4n+2)$ can be prospective GUFs. The simplest among these is
a spinor of $SO(14)$. But, with a trivial style of breaking
$SO(14)$ down to $SO(10)$ leads to two left-handed {\bf 16} and
two right-handed {\bf 16}, leaving null families at low energy.
Therefore, to obtain chiral families one must twist the group so
that $Q_{em}$ is not of the standard type. One attractive
example\cite{kimsu7} is the $SU(7)$ subgroup of $SO(14)$ with the
representation, [3]+[$\bar 2$]+[1].  It comes from a spinor of
SO(14). We use $SO(14)$ and $SU(7)$ interchangeably up to a
singlet, ($A=1,2,\cdots,7$)
\begin{equation}  \label{spinor}
{\bf 64}=\psi^{ABC}+\psi_{AB}+\psi^A+{\bf 1}=35\oplus \overline{
25}\oplus 7\oplus 1
\end{equation}
With  $Q_{em} = {\rm diag.}(-\frac13,-\frac13,-\frac13,1,0,1,-1)$
which is twisted compared to the standard one, the left-handed
lepton doublets are the same but the right-handed lepton doublets
are electromagnetically shifted by +1 unit for one family and --1
for the other family,
$$
\left(\matrix{\nu_e\cr e\cr}\right)_L,\ \ \left(\matrix{\nu_\mu\cr
\mu\cr} \right)_L,\ \ \left(\matrix{\tau^c\cr
\nu^c_\tau\cr}\right)_R,\ \ \left(\matrix{E^-\cr E^{--}\cr}
\right)_R
$$
Thus, it explains three families of leptons successfully, but for
quarks it fails because the $b$ quark is known to decay by the CKM
structure\cite{kane}. Since 1984, the GUF idea has been replaced
by the heterotic string group $E_8\times E_8^\prime$ since $E_8$
is big enough to house the three families. Even though $E_8$
allows only real representations, the extra dimensions can work as
twisting the group, leading to chiral families at low energy
\cite{candelas,dixon,iknq}.\\

\noindent {\bf Field theoretic orbifold compactification(FTOC)}--
In the early string days the compactification of the extra
dimensions was done with the strings located in the extra
dimensions. In recent years, however, field theories in extra
dimensions opened up new possibilities and subsequently field
theoretic orbifolds have been used for breaking the GUT
group\cite{kawamura}. This FTOC is extended to break the GUF, and
we intend to discuss the family unification in extra dimensions
with orbifold compactification.

In the $Z_2\times Z_2^\prime$ compactification, the massless mode
comes from the sector with $(Z_2,Z_2^\prime)=(+,+)$. The others
are heavy with mass at the inverse compactification scale.
 A spinor of $SO(14)$ breaks down to
\begin{equation} \label{bulk}
\Psi^{ABC} \oplus \Psi_{AB} \oplus \Psi^{A} \oplus \Psi= {\bf
16}\otimes {\two}_F \oplus \overline{\bf 16} \otimes {\two}'\,,
\end{equation}
where the RHS is the decomposition into $SO(10)\times SU(2)\times
SU(2)'$ and the anti-symmetrization of the indices are assumed.
Since we are dealing with $SO(4n+2)$ groups, the models considered
do not have the anomaly problem.

A 5D $SO(14)$ spinor has four left-handed and four right-handed 4D
$SO(10)$ spinors. Under the torus compacification, these eight
$SO(10)$ spinors form four massive Dirac spinors and are removed
from the low energy spectrum. But twisting can allow some zero
modes. Let the $Z_2$ action makes the right-handed component of a
5D spinor heavy (breaking one supersymmetry if there was). In
other words, only 4 left-handed $SO(10)$ spinors(one left-handed
$SU(7)$ spinor) in 4D remain as zero modes. It is represented
under $SU(5)\times SU(2)\times U(1)$ as:
\begin{equation}
\begin{array}{llllllllll}
\Psi^{ABC} &=& \psi^{\alpha\beta\gamma} & {(\tenb,\one)}_6 &
\oplus
    & \psi^{\alpha\beta i} & {(\ten,\two)}_{-1}    & \oplus
    & \psi^{\alpha ij} & {(\five,\one)}_{-8}    \\
\Psi_{AB} &=& \psi_{\alpha\beta} & {(\tenb,\one)}_{-4} & \oplus
    & \psi_{\alpha i} & {(\fiveb, \two)}_3 & \oplus
    & \psi_{ij} & {(\one,\one)}_{10} \\
\Psi^{A} &=& \psi^{\alpha} & {(\five, \one)}_2 & \oplus
    & \psi^{i} & {(\one,\two)}_{-5} & & &
\end{array}
\end{equation}
where the total number of $\ten$ and $\tenb$ is four which is the
number of massless $SO(10)$ spinor zero modes. Here, the upper
case Roman letters $A,B,C,\cdots$ are the $SU(7)$
indices($1,2,\cdots,7$), the lower case Greek letters
$\alpha,\beta,\gamma, \cdots$ are the $SU(5)$
indices($3,4,\cdots,7$), and the lower case Roman letters $i,j$
are the $SU(2)_F$ indices 1, 2. We can assign $\lambda' = -1$ to
the $Z_2'$ parity  of the whole $SU(7)$ spinor ($\Psi^{ABC},
\Psi_{AB}, \Psi^A$), leaving the following zero modes
\begin{equation}
\label{zeromode} {(\ten,\two)}_{-1}\,,\quad
{(\fiveb,\two)}_{3}\,,\quad {(\one,\two)}_{-5}\,,
\end{equation}
which is exactly the anomaly free combination of the flipped
$SU(5)$ model\cite{su51} after orbifold compactification. Thus,
this consistent choice of $Z_2'$ parity picks up one irreducible
representation of ${\bf 16}\otimes \two$ of $SO(10)\times SU(2)$
in 4D among the full spinor of $SO(14)$ shown in Eq.(\ref{bulk}).
But where is the third family? Here comes the short-coming of
$SO(14)$. We must introduce another spinor of $SO(14)$, which may
be justified in string models. Note that Kawamura used the FTOC as
a solution of the split multiplet problem. But we used the FTOC
for twisting the space-time and the gauge group to obtain multi
families\cite{barr,hwang}.\\

\noindent{\bf Higgs content}-- Because it is the flipped $SU(5)$,
it is possible to break the GUT group without the adjoint
representation for the Higgs sector. $\langle {\bf
\overline{10}}_{+1}\rangle$, containing a neutral component, can
break the $SU(5)\times U(1)$ down to the standard model gauge
group at a GUT scale. Thus, for the GUT breaking, we need another
${\bf \overline{16}}$ of $SO(10)$ for a Higgs field. Since it is a
representation of the broken group say $SO(10)$, we put it at the
orbifold fixed point where the GUT group is broken, say A. As we
show below, most of $\overline{\bf 16}$ are removed, combining
with one ${\bf 16}$ from the additional ${\bf 64}$ of $SO(14)$.

What are the electroweak Higgs for the $SU(2)\times U(1)$
breaking? ${\bf \overline{5}}$ and ${\bf 5}$, appearing in ${\bf
16}$ and $\overline{\bf 16}$ cannot be these electroweak scale
Higgs since the electroweak Higgs should carry the flipped $U(1)$
quantum number 2 and --2, i.e. ${\bf 5}_2$ and $\overline{\bf
5}_{-2}$. These electroweak Higgs must be a vector representation
${\bf 10}$
of $SO(10)$, and can be put at the orbifold fixed point A.\\

\noindent{\bf Missing partner mechanism for doublet-triplet
splitting and needed Higgs fields}-- We introduced two
$SU(2)_F$--doublet spinors of $SU(7)$. For the Higgs fields, let
us introduce ${\five}_2$ and ${\fiveb}_{-2}$ in the bulk, and in
addition \{${\tenb}_1\oplus {\five}_{-3} \oplus {\one}_5$\} at the
asymmetric fixed point, which are $SU(2)_F$--singlets. Toward a
detail discussion on the mass matrices of light fermions and the
doublet-triplet splitting mechanism, let us name two
$SU(2)_F$--doublets of $SO(14)$ spinor as
\begin{equation} \label{mdoub}
T_i({\ten}_{-1}), {\fb}_i({\fiveb}_3), E_i^c({\one}_{-5}), \quad
\mbox{\rm and} \quad T'_i({\ten}_{-1}), {\fb'}_i({\fiveb}_3),
E_i^{'c}({\one}_{-5}) ,
\end{equation}
where the family indices $i=1,2$ and $SU(2)_F$--singlets as
\begin{equation} \label{msing}
\Hb(\tenb_{1}), h'({\five}_{-3}), \varphi({\one}_{5}), \quad
\mbox{\rm and} \quad h({\five}_2), \hb({\fiveb}_{-2})\,,
\end{equation}
and the components of each multiplet as
\begin{equation}
{\ten}_{-1} : \pmatrix{d^c & q \cr q & \nu^c} \quad\quad
{\fiveb}_{3} : \pmatrix{u^c \cr \ell} \quad\quad {\fiveb}_{-2} :
\pmatrix{\overline{D} \cr h^+ } \quad\quad {\five}_{+2} :
\pmatrix{D \cr h^- } \,
\end{equation}
where $\overline{D}$ and $h^+$ carries the hypercharge $1/3$ and
$1/2$, respectively.

It was shown that  $SU(2)_F$--doublet fields $\{\chi_i^1,
\chi_i^2\}= 2{(\one, \two)}_0$ are needed at the asymmetric fixed
point to obtain the needed interactions\cite{hwang}. The following
two discrete symmetries
\begin{equation} \label{disc}
Z_2^\chi\ :\ \chi^1\rightarrow -\chi^1\ ,\ \ \varphi\rightarrow
-\varphi, \qquad \quad Z_2^H\ :\ \Hb\rightarrow -\Hb
\end{equation}
while the other fields are invariant under $Z_2^\chi$ and $Z_2^H$,
guarantees the following superpotential,
\begin{equation} \label{wh}
W_{H}=\Hb\Hb\hb + T'T'h + T'\fb'\hb + \fb' E^{\prime c} h +
\fb'h'\chi^2 + E^{\prime c}\varphi\chi^1.
\end{equation}
 We do not allow $h\hb$ term in
the superpotential, which is anticipated in the superstring
models. By the development of VEV along the $D$-flat(and $F$-flat)
direction $(T' \Hb \chi^1)(\chi^1\chi^2)$,
\begin{equation}
\langle\nu^c_{T'_1} \rangle = \langle\overline{\nu}^c_{\Hb}
\rangle = \frac{1}{\sqrt{2}}\langle\chi_2^1\rangle
=\langle\chi_1^2\rangle= M_{G},\label{symm}
\end{equation}
both $q_{T'_1}$ and $q_{\overline{H}}$ are either eaten by the
heavy gauge bosons or made heavy by the supersymmetric Higgs
mechanism. From the superpotential terms in Eq.(\ref{wh}) the
components $d^c_{T'_2}, D_h, \overline{d}^c_\Hb, \overline{D}_\hb,
\fb'_2,E_1^{\prime c}$ and one linear combination of $h^+$ and
$h'$ become massive after the symmetry breaking, while $h^-$ and
the other linear combination of $h^+$ and $h^\prime$ remain
massless and fulfil the doublet-triplet splitting. The rest
massless components \{$d^c_{T'_1}, q_{T'_2}, u^c_{F'_1},
\ell_{F'_1}, E^c_2$\} form the third generation family.

Furthermore, with suitable assumptions on the scales of vacuum
expectation values and additional terms in the superpotential, we
obtain mass matrices of the form,
\begin{equation}
{M^{u,d}\over M^{u,d}_{33}} \approx
  \pmatrix { 0 & \epsilon' & 0 \cr
 \epsilon' & \epsilon & \epsilon \cr \epsilon & 0 & 1}\,,
\qquad\quad {M^{e}\over M^{e}_{33}} \approx
  \pmatrix { 0 & \epsilon' & \epsilon \cr
 \epsilon' & \epsilon & 0 \cr 0 & \epsilon & 1}\,.
\end{equation}
which gives a qualitatively correct mass spectrum and CKM mixing
matrix elements\cite{hwang}.\\

\noindent{\bf Superstring connection}-- The FTOC is an easy way to
construct an effective 4D theory. The field contents in 5D we
considered are better to arise from a string compactification. In
this regard, a $Z_2$ orbifold compactification in 6D is
worthwhile. In one compactification\cite{kangsin}, we obtained a
supersymmetric model with $SO(16)$ gauge group and matter
hypermultiplets
\begin{equation}
{\bf 128}\oplus \ {\rm sixteen}\ {\bf 16}
\end{equation}
which form an anomaly free 6D model. By torus compactification and
orbifolding, we can obtain a flavor unification in 4D\cite{kangsin}.

\section*{Acknowledgments}
This work is supported in part by the BK21 program of Ministry of
Education, and the KOSEF Sundo Grant.

\end{document}